\documentstyle{mn}
\title{Large-scale distribution of quasars in deep pencil-beam surveys}

\author[Andrey V. Kravtsov]
       {Andrey V. Kravtsov \\
Astronomy Department, New Mexico State University, Box 30001, Department 4500,
         Las Cruces, NM 88003-0001, USA}

\begin{document}
\maketitle
\begin{abstract}

We have used integral two-point spatial correlation function and its 
second derivative  to analyze the distribution of quasars in 
three very deep surveys published in the literature. 
Statistically significant ($\sim2-3\sigma $) correlations were found at 
scales of $\sim 50-100h^{-1}$ Mpc
in all of the analyzed surveys. We have used the
friend-of-friend cluster analysis to show that these correlations can be 
explained by the presence of relatively small quasar clusters
($3-6$ objects) which may possibly belong to larger structures such 
as Large Quasar Groups found in the bigger surveys. The sizes of these 
clusters along the redshift direction and distances between them 
are similar to those for structures 
found recently in studies of CIV absorption systems.
These results present further evidence for the existence of large-scale 
structures at redshifts $z\sim 1-2$. 

\end{abstract}
\section{Introduction}
 Quasi-stellar objects (QSOs) have proved to be very useful cosmological
probes of the high-redshift Universe. The availability of large homogeneous 
surveys has made it possible to study the spatial distribution of quasars
using statistical tools such as the two-point spatial correlation function. 
These studies have revealed that quasars are strongly 
clustered at scales of $r<20h^{-1}$ Mpc\footnote{All the quoted scales are 
comoving and were calculated assuming flat
Friedman cosmology with $H_0=100h$ km/s/Mpc and $\Lambda=0$.}
(e.g. Iovino et al. 1991; 
Andreani \& Cristiani 1992; Mo \& Fang 1993; Komberg et al. 1994; 
Shanks \& Boyle 1994 and references therein). Although there is still an 
ongoing discussion about whether 
these correlations are evolving with redshift, the fact 
that the amplitude and shape of the quasar correlation function are
roughly similar to those of low-redshift objects 
(e.g. Andreani \& Cristiani 1992; Mo \& Fang 1993; Komberg et al. 1994; 
Shanks \& Boyle 1994) shows that
quasars may possibly be used as tracers of the matter distribution at medium 
 redshifts ($z\sim1-2$). 
 
In addition to the statistical results, evidence has been found for 
structures in the quasar distribution at even larger ($\sim 100-150h^{-1}$
Mpc) scales (Webster 1982; Crampton, Cowley \& Hartwick 1987, 
1989; Clowes \& Campusano 1991a, 1991b; Graham, Clowes \& Campusano 1995;
Komberg et al. 1996). These structures, which we call Large 
Quasar Groups (LQGs), consist of $\sim 10-25$ quasars and have sizes and 
density contrasts similar to those of nearby superclusters of galaxies 
(Komberg et al. 1996).
Recently, similar high-redshift structures have been found in the distribution 
of absorbers (e.g. Jakobsen et al. 1986; 
Sargent \& Steidel 1987; Jakobsen \& Perryman 1992; Dinshaw \& Impey 1996; 
Williger et al. 1996)
This indicates that supercluster type structures are observed at 
redshifts $\sim 0.5-2.5$ -- the fact which may be used as a useful 
constraint for existing models of structure formation and evolution. 

If the quasar large-scale structures are real, we should be able to detect 
them using statistical methods. Deng et al. (1994) used the second derivative 
of the integral two-point spatial correlation function to search
for typical scales in the distribution of quasars. They argued
that there exists a typical scale of about $95h^{-1}$ Mpc and that 
this scale may be related to the specific shape of the initial perturbation
spectrum. 
In this paper we present a study of the quasar distribution in 
three deep surveys. We have used statistical methods similar to those of 
Deng et al. (1994). Our goals and strategies, however, were somewhat 
different. First, we have chosen to study only the deepest quasar surveys.
The number density of quasars in these surveys is much higher than in the
larger surveys which makes them more sensitive to the presence of structures
at scales $50-100h^{-1}$ Mpc. Second, we have planned to look for 
evidence of individual large-scale structures rather than
for typical scales in the overall quasar distribution (the numbers of objects
in the deep samples
that we used are probably too small for that). 

The plan of the paper is as follows: in 
Section 2 we present the quasar surveys and briefly discuss 
the limitations they impose, in Section 3 we describe the statistical 
methods; in Section 4 we present 
the results and discuss the implications in Section 5. 
\section{Quasar samples}
The relatively low number density of quasars in existing {\em bright}
quasar surveys  makes some of them 
insensitive to spatial correlations at scales $< 100h^{-1}${\ }Mpc
(Komberg et al. 1994). On the other hand, the deepest quasar surveys 
usually have high enough quasar densities (mean separation between objects
$\sim 20-40h^{-1}$ Mpc) to be sensitive to the presence of supercluster 
size ($\sim 100h^{-1}$ Mpc) structures. These surveys are usually 
composed of one or several ``beams'' sampling small areas on the sky.
The beam sizes perpendicular to the line of sight
($\sim30-50h^{-1}$ Mpc) are much smaller  
than the sampled distances  along the line of sight ($\sim2000h^{-1}$ Mpc). 
The ``one-dimensional'' nature of these surveys makes analysis and 
interpretation of the results relatively simple because if significant 
correlations exist, we can then look directly 
for the structures responsible for them. Such analysis is difficult for larger
surveys because their sky geometry is usually quite complicated.
However, the presently available deep surveys have small numbers of quasars
(usually few tens) in a beam, which complicates and limits any statistical
analysis. One then must use statistics suitable for the analysis of 
small data sets. 

For our study we have chosen three quasar surveys published in the 
literature: i) a survey  by Koo \& Kron 1988 (hereafter KK survey),
which was compiled using $UBVI$ photometry, variability and proper motions 
of candidates; ii) a grism survey by Zitelli et al. 1992 
(hereafter (ZM)$^2$B); 
and iii) a deep survey by Boyle, Jones \& Shanks 1991 (hereafter BJS) 
in which quasar 
selection was based on multicolor photometry of the candidates.
The BJS survey is composed of three beams covering approximately the 
same area on the sky ($\approx 0.3$ sq.deg.). We have, therefore, analyzed
these three beams (BJS1, BJS2, BJS3) separately. Also, following (ZM)$^2$B,
their
survey can be split into two samples with different levels of completeness: 
a sample of
24 quasars complete to $J=20.85$ in an area of 0.69 square degrees; and 
sample of 28 quasars complete for $20.85\leq J\leq 22.0$ in a circular 
area of 0.35
square degrees which is contained in the previous one. 
In our analysis we have used a sample of 38 quasars
complete for $J\leq 22.0$\footnote{In this sample we have included all quasars
inside the 20-arcmin circle defined in (ZM)$^2$B.}.
 The basic characteristics of all of our samples  -- 
surveyed area in square degrees, limiting magnitude, number of quasars, 
and redshift interval -- are presented in Table 1. 
\begin{table}
  \caption{Surveys used in the analysis.}
  \begin{tabular}{@{}ccccc}
\hline 
    Sample        &  Area             &  Limiting    & Number of & Redshift   \\
                  & (sq.deg.)         &  magnitude   & quasars   & interval   \\
\hline 
    KK            &       0.29        & $B < 22.5$   &    28     &  $0.9-3.2$ \\
  (ZM)$^2$B       &       0.35        & $J < 22.0$   &    38     &  $0.4-2.8$ \\
    BJS1          &       0.29        & $b_j < 21.8$ &    21     &  $0.6-2.9$ \\
    BJS2          &       0.33        & $b_j < 22.0$ &    19     &  $0.6-2.9$ \\
    BJS3          &       0.27        & $b_j < 21.8$ &    20     &  $0.6-2.9$ \\
\hline
\end{tabular}
\end{table}
%
\section{Method}
%
In this study, we have used the integral two-point correlation function 
(e.g. Mo et al. 1992; Mo \& Fang 1993) which is related to the usual 
differential correlation function $\xi(r)$ as follows:
$$\overline{\xi}(r)=\frac{3}{r^3}\int^r_0x^2\xi(x)dx.$$ 
This choice was made because the integral correlation function 
is more stable than $\xi(r)$ in the analysis of small data sets.
This is important because we intended to look for features in the 
correlation function 
at large scales, where signal-to-noise ratio is usually small. 
We estimate $\overline{\xi}(r)$ using a standard estimator 
(e.g. Mo \& Fang 1993):
$$\overline{\xi}(r)=\frac{\Pi_{obs}(r)}{\Pi_{rnd}(r)}-1,$$
where $\Pi(r)$ is the number of pairs with separations less than 
$r$ in an analyzed sample, while $\Pi_{rnd}(r)$ is the corresponding number 
of pairs averaged over an ensemble of random samples. 
To estimate $\Pi_{rnd}(r)$, a thousand 
random catalogs were created using the {\em smoothing} method
of Mo \& Fang (1993). 
In this method, random samples are constructed by assigning 
each object a random position in the sky within the boundaries of 
the sample and by drawing redshifts
randomly from a smoothed version of the original redshift distribution. 
The smoothed redshift distribution is obtained by 
averaging the number of quasars in the redshift interval $\Delta z=0.6$
(corresponding to $\sim 500h^{-1}$ Mpc at $z\approx 1.5$)
around a given redshift. This width of the interval was chosen so that 
it is small enough to preserve the overall survey selection envelope 
but is considerably larger than the scales we are interested in 
($\sim 100h^{-1}{\ }$ Mpc).

The whole analysis is similar to that of Mo et al. (1992). It is outlined 
in the following series of steps.
\begin{enumerate}[(iii)]
\item First, we compute the function $\Xi(r)=\overline{\xi}(r)+1$;
\item Then we find the slopes $T_1(r)$ and $T_2(r)$ (or corresponding angles
      $\Theta_1$ and $\Theta_2$) of $\Xi(r)$ for every $r$ bin using 
      linear regression of the relation $\log \Xi(r)-\log r$ in the intervals 
      $[r-\Delta r,r]$ and $[r,r+\Delta r]$, where $\Delta r$ is 
      the {\em smoothing scale}. 
      Smoothing with a given $\Delta r$ effectively damps out 
      amplitude fluctuations in $\overline{\xi}(r)$ on scales $r<\Delta r$. 
      The choice of $\Delta r$ is determined by the mean separation 
      $\overline{r}$
      between objects in a sample -- at scales smaller than $\overline{r}$ 
      the shot noise is significant and must be suppressed. 
\item Finally, we construct the function 
      $\Delta \Theta(r)=\Theta_2(r)-\Theta_1(r)$,
      i.e. the smoothed second derivative of $\overline{\xi}(r)$. 
      Significant 
      changes in the slope of $\log\Xi(r)-\log r$ (changes in the shape 
      of $\overline{\xi}(r)$) reflect the inhomogeneities in the distribution 
      of objects at the corresponding scales. These changes result in sharp peaks 
      in $\Delta \Theta(r)$ with scale independent amplitude 
      (Mo et al. 1992; Deng et al. 1994). This is 
      very useful when working on the large scales where the amplitude of the 
      correlation function is small.       
\end{enumerate}

We have estimated rms errors in $\overline{\xi}(r)$ using sample-to-sample 
deviations in the random catalogs as well as using analytical formula presented by
Mo et al. (1992). We have constructed $n=1000$ random catalogs 
using the same smoothing procedure described above which we then 
analyzed in the 
same way as real data. In this way we could estimate the fluctuation 
in the number of 
pairs (and thus the fluctuation of $\overline{\xi}(r)$) at a given scale as:
$$\Delta \Pi=\sqrt{\frac{\sum_{i=1}^n(\Pi_i - \langle \Pi\rangle )^2}{n-1}},$$
here $\langle \Pi\rangle$ is the number of pairs averaged 
over all random catalogs,
and $\Pi_i$ is the corresponding number of pairs in $i$-th random catalog.
The deviation of $\Delta\Theta(r)$ was estimated in a similar way:
$$\sigma_{\Delta\Theta}=\sqrt{\frac{\sum_{i=1}^n(\Delta\Theta_i - 
\langle\Delta\Theta\rangle)^2}{n-1}}.$$
These estimates were then compared with the values given by the 
theoretical formulae (Mo et al. 1992). For one-dimensional samples they are:
$$N_{r\pm\Delta r}=\frac{N}{2}\left(\frac{N}{L}\Delta r\right),{\ \ \ }
\sigma_{\Delta\Theta}=\left(\frac{2L}{\Delta r}\right)^{\frac{1}{2}}
\frac{1}{N},$$
where $L$ is the sample extent in the redshift direction, 
$N$ is the number of objects in the sample, and $\Delta r$ is the smoothing 
length adopted in the construction of $\Delta\Theta (r)$.
We have found that errors predicted by these formulae agree well with 
values derived from the sample-to-sample variations. We have, therefore, used
these formulae to estimate the significance of peaks in $\Delta\Theta(r)$.

Finally, we have used {\em friend-of-friend} cluster analysis 
(e.g. Einasto et al. 1984) to study the quasar distribution in the 
samples directly. The details of the cluster identification procedure and the 
way of estimating the probability for a cluster to be random 
are described in Komberg et al. (1996). 

\section{Results}

The statistics described in the previous section were computed for all five
quasar samples. The functions $\Xi(r)=\overline{\xi}(r)+1$ and 
$\Delta \Theta(r)$ for each sample are plotted in figs. 1 and 2. On these
plots the functions are shown by solid lines and one sigma error
envelope by the thin dashed lines. The features in the correlation 
function which we discuss below are indicated by arrows. 
The mean separations between 
quasars in the samples are $\sim20-30h^{-1}$ Mpc for the KK and (ZM)$^2$B
surveys, and $\sim40h^{-1}$ Mpc for the BJS survey. This 
determined the choice of smoothing length for $\Delta \Theta(r)$ -- 
$20h^{-1}$ Mpc and $40h^{-1}$ Mpc correspondingly (on the plots 
$\Delta \Theta(r)=0$ at $r<\Delta r$). The integral correlation function
was computed with a bin size of $2.5h^{-1}$ Mpc for the KK and 
(ZM)$^2$B samples and with a $5h^{-1}$ Mpc bin for the BJS samples.
Below we describe results for each sample separately.
\subsection{KK}
Quasars in the KK sample show strong correlations at small scales 
$r<10h^{-1}$ Mpc (fig. 1a). Statistical significance of the correlation 
signal is comparable to that for samples containing ten times more quasars
(e.g. Mo \& Fang 1994; Boyle \& Shanks 1994).
While the number of objects in this sample is quite small (28), the 
high number density of quasars assures relatively good statistics of close 
pairs (see Komberg et al. 1994 for discussion of close quasar pair statistics).
The same is true for the (ZM)$^2$B sample. 

The most interesting feature of the correlation function is a ``bump'' at
$\sim 50h^{-1}$ Mpc which is also seen as a sharp negative peak in 
$\Delta\Theta(r)$ (fig. 1b). The significance of the peak is 
$\sim 2.5\sigma$. Using friend-of-friend cluster analysis
we have found a quartet and two triplets (i.e. 10 quasars out of a total 
of 28) 
with sizes along the redshift direction of $43$, $30$, and $34h^{-1}$
Mpc correspondingly. The estimated 
probability for these clusters to be random is quite small for the quartet 
and one of the triplets ($0.01$ and $0.08$) while for the second triplet
it is rather
high ($\sim 0.5$). However, the method we use to estimate this probability 
is not 
very reliable for triplets (and does not work at all for pairs) because 
the uncertainty of density within the triplet is high. We have found that 
the excess of pairs at separations $r\sim 30-50h^{-1}$ Mpc (causing 
the bump in $\overline{\xi}(r)$) is due to the presence of these small 
clusters.
\subsection{(ZM)$^2$B}
For this sample the correlation signal at small separations is 
also quite high. The ``bump'' is present at 
$\sim 110h^{-1}$ Mpc (fig. 2a)
with a corresponding peak in $\Delta\Theta(r)$ (fig.2b) which is significant
at the $\sim 2.5\sigma$ level. Applying the cluster analysis,
we have found three quartets and two quintets (22 quasars out of a total of 
38). 
One of the quartets is probably random
(estimated probability is $\sim 0.6$), the probability to be random for 
the rest of the clusters is small ($<0.1$) -- their sizes in $z$-direction 
lie in the range $25-50h^{-1}$ Mpc.  
Four of these clusters form two ``pairs'' $100\pm 20h^{-1}$ Mpc creating 
the excess of quasar pairs at these separations.
\subsection{BJS1}
The statistics for the first BJS sample are shown in figs. 2a and b. 
There is 
a rise in $\Xi(r)$ at a separation of $\sim40h^{-1}$ Mpc. This rise 
is not reflected in $\Delta\Theta(r)$ because its scale is equal to the 
smoothing scale. Using cluster analysis we have detected two clusters 
in this sample -- a triplet and a quintet (8 out of a total of 21 quasars). 
The $z$-sizes of both clusters are $\sim 35h^{-1}$ Mpc. The estimated 
probability to be random is $0.05$ and $0.01$, respectively. Both 
of these clusters contribute to the excess of quasar pairs at separations
of $30-40h^{-1}$ Mpc. 
\subsection{BJS2}
Results for the BJS2 sample are shown in figs. 2c and d. The decay of 
$\Xi(r)$ at $r<30h^{-1}$ Mpc is caused by the fact that the number density 
of quasars in this sample is considerably lower than in KK and (ZM)$^2$B
and even lower than in BJS1 and BJS3.
This sample therefore just lacks close pairs. The correlation function
fluctuates considerably at scales $\sim 150-200h^{-1}$ Mpc (there is 
a significant excess of pairs at these separations as compared to the 
random distribution). These fluctuations correspond to the positive 
and negative peaks in $\Delta\Theta(r)$ at $\sim 150h^{-1}$ Mpc and 
$\sim 180h^{-1}$ Mpc. Using cluster analysis we have found 
a triplet ($z$-size $\sim 20h^{-1}$ Mpc), a quartet 
($z$-size $\sim 20h^{-1}$ Mpc), and a sextet 
($z$-size $\sim 75h^{-1}$ Mpc). The probability that each individual cluster
is random is smaller than $0.05$. Our analysis has shown that the 
excess of pairs
at separations $150-200h^{-1}$ Mpc is explained by the distance between
the quartet and the sextet ($\sim 180h^{-1}$ Mpc). 
\subsection{BJS3}
The correlation function for this sample (fig. 2e) has two broad ``bumps''
at scales $\sim 100h^{-1}$ Mpc and $\sim 200h^{-1}$ Mpc. There are
two negative peaks in the $\Delta\Theta(r)$ corresponding to these  
fluctuations
(fig. 2f). Statistical significance of both peaks is $\sim 2\sigma$. The 
cluster analysis failed to detect any large clusters in this sample. 
However, the distribution of quasars in this sample is quite interesting.
There are seven relatively close pairs (distances between quasars in
5 of them are 
less than $30h^{-1}$ Mpc and in the other two $\sim 40h^{-1}$ Mpc) 
separated {\em from each other} by either $80-100h^{-1}$ Mpc or 
$180-200h^{-1}$ Mpc. This causes the fluctuations in $\Xi(r)$ 
at the corresponding scales. 
\section{Discussion and conclusions}
The results presented in the previous section suggest that the distribution
of quasars in the analyzed samples is not homogeneous at scales of a few 
tens of megaparsecs. Many quasars belong to clumps of sizes $30-70h^{-1}$
Mpc. The clumps are often separated by $100-200h^{-1}$ Mpc, which creates
a pair excess at the corresponding scales. 
Qualitatively this quasar distribution is very similar to that 
of CIV absorption systems discussed in the recent paper by 
Williger et al. (1996). They have found that their CIV sample 
contains two groups of $7$ and $5$ absorbers of sizes 
$\sim43\times17\times69h^{-3}$ Mpc$^3$ and 
$\sim25\times4\times53h^{-3}$ Mpc$^3$
(comoving) located at $z\sim2.3$ and $z\sim2.5$, respectively. 
The distance between these two groups ($\sim 50-120h^{-1}$ Mpc) results
in ``beating'' (the pair excess) giving rise to the correlation
signal at these separations ($3.5\sigma$ significance level). A number
of smaller clumps were also detected. The similar clusters of CIV absorbers
were also found in earlier studies by Jakobsen \& Perryman (1992), 
Foltz et al. (1993), and Dinshaw \& Impey (1996). 
Recently, Lespine \& Petitjean (1996) presented
evidences for a coherent structure extended over $\sim80h^{-1}$ Mpc at 
$z\approx2$ in the distribution of metal absorption systems. 
Although the numbers of CIV systems 
are also small, the similarity of the results may suggest that 
both quasars and CIV absorbers may trace the same kind of underlying 
structures in the matter distribution. 
Unlike Deng et al. (1994) 
we did not observe any evidence for a periodic signal in the function
$\Delta\Theta(r)$\footnote{Although $\Delta\Theta(r)$ seems to 
be periodic,
most of the peaks are not significant (their height is $<1\sigma$).}. 
This may be caused by the small number statistics. 

The clumpy distribution of quasars in the analyzed samples 
is consistent with the recent studies of the quasar distribution in the
larger samples (Crampton, Cowley \& Hartwick 1987, 
1989; Clowes \& Campusano 1991a, 1991b; Graham, Clowes \& Campusano 1995;
Komberg et al. 1996). These samples were found to contain several 
relatively rich ($\sim10-25$ QSOs) groups of quasars with sizes in 
the redshift direction of $\sim70-160h^{-1}$ Mpc. The small extent of the
pencil-beam samples perpendicular to the line of sight 
prevents detection of such large groups. However, the detected smaller 
clumps can easily be parts of larger systems. It would be very interesting
to check this by studying larger deep samples which are currently 
underway (e.g. Hall et al. 1996). 

If quasars and CIV absorption systems trace the matter distribution at 
high redshifts as galaxies or galaxy clusters do at low redshifts, 
their clumpy distribution suggests that large-scale inhomogeneities
similar to the nearby superclusters were already distinct at $z\sim1-2$. 
This information may provide some useful insights into the physics 
of high redshift Universe. The fact that we see structures at 
redshifts $z\sim1-2$ similar to the superclusters at $z\sim0$ 
(Komberg et al. 1996), for instance,
favors low-density $\Lambda$CDM or low-density CDM models in which 
perturbation amplitude at large scales stops growing at $z\geq 1$.
On the other hand, rather high quasar-quasar correlations at small 
separations and high number density 
contrasts in the detected quasar groups
may indicate that the distribution of quasars is highly biased with respect
to the matter distribution. 
Although present available surveys are too small to provide 
a statistically reliable estimate of the power spectrum $P(k)$, 
in the future, 
with bigger quasar samples and better models for both QSOs and CIV absorbers, 
we will be able to get useful constraints on the spectrum, and thus on 
the theories of structure formation, for a wide range of scales and redshifts
(e.g. Komberg \& Lukash 1994). 
\section{Acknowledgments}
I would like to thank B.V.Komberg and V.N.Lukash for many fruitful
discussions. Invaluable assistance in improving the presentation of
the paper was provided by Matthew Carlson. 
At different stages this project was supported by an ESO C\&EE grant 
(No. A$-01-152$), the 
Russian Foundation for Fundamental Research (project $93-02-2929$), 
and the International Science Foundation (project code MEZ300).

\pagebreak
\begin{center}
FIGURE CAPTIONS
\end{center}

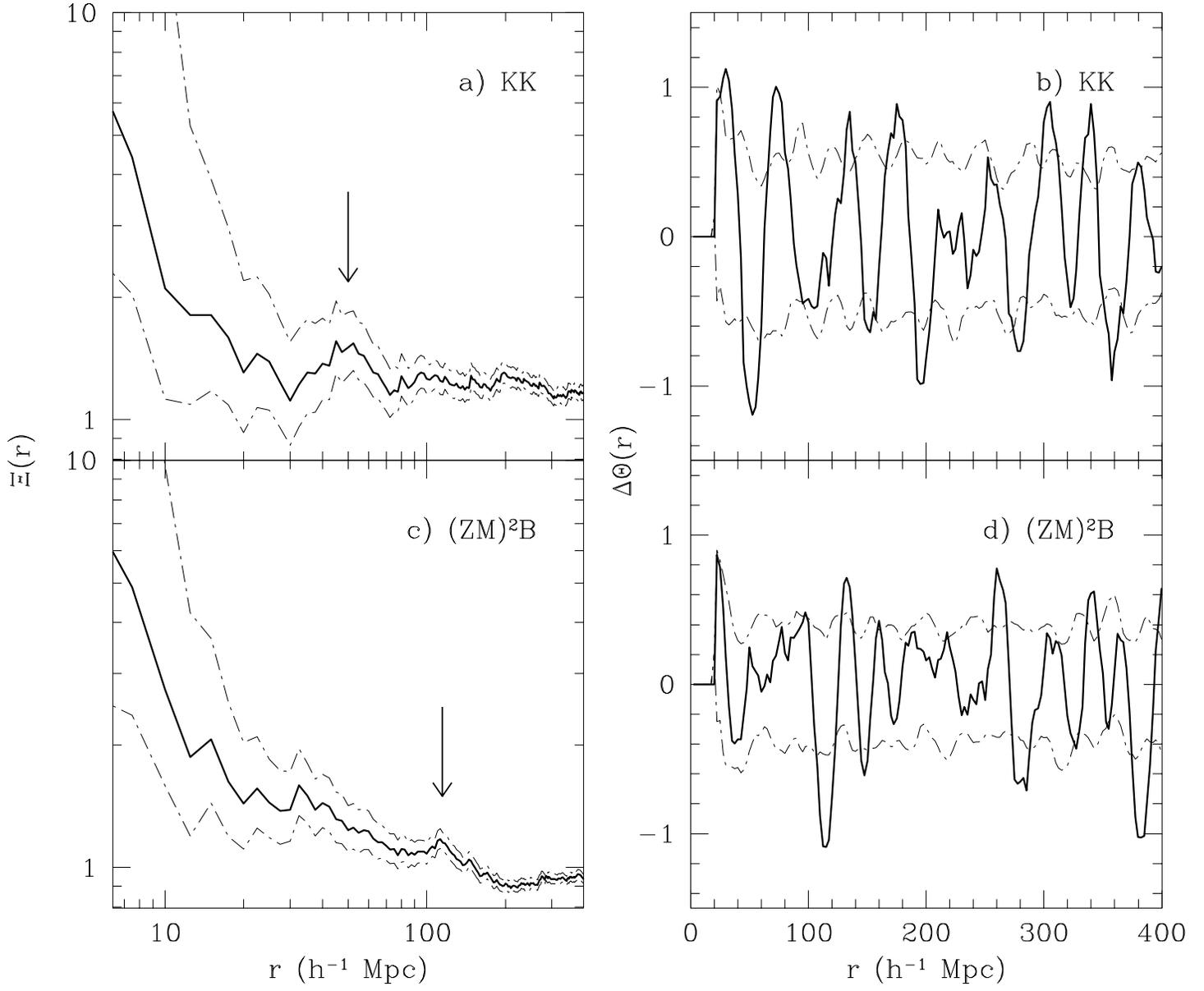
{\bf Figure 1.} The integral two-point spatial correlation function
$\overline{\xi}(r)$ for the KK (panel a) and (ZM)$^2$B (panel c) samples
and its smoothed second derivative (panels b and d) $\Delta\Theta(r)$ 
(see definitions in Section 3). The functions are solid lines. 
The $1\sigma$ 
error envelope is shown by thin dashed lines. The features in the 
correlation function discussed in the text (Section 4) are indicated 
by arrows. 

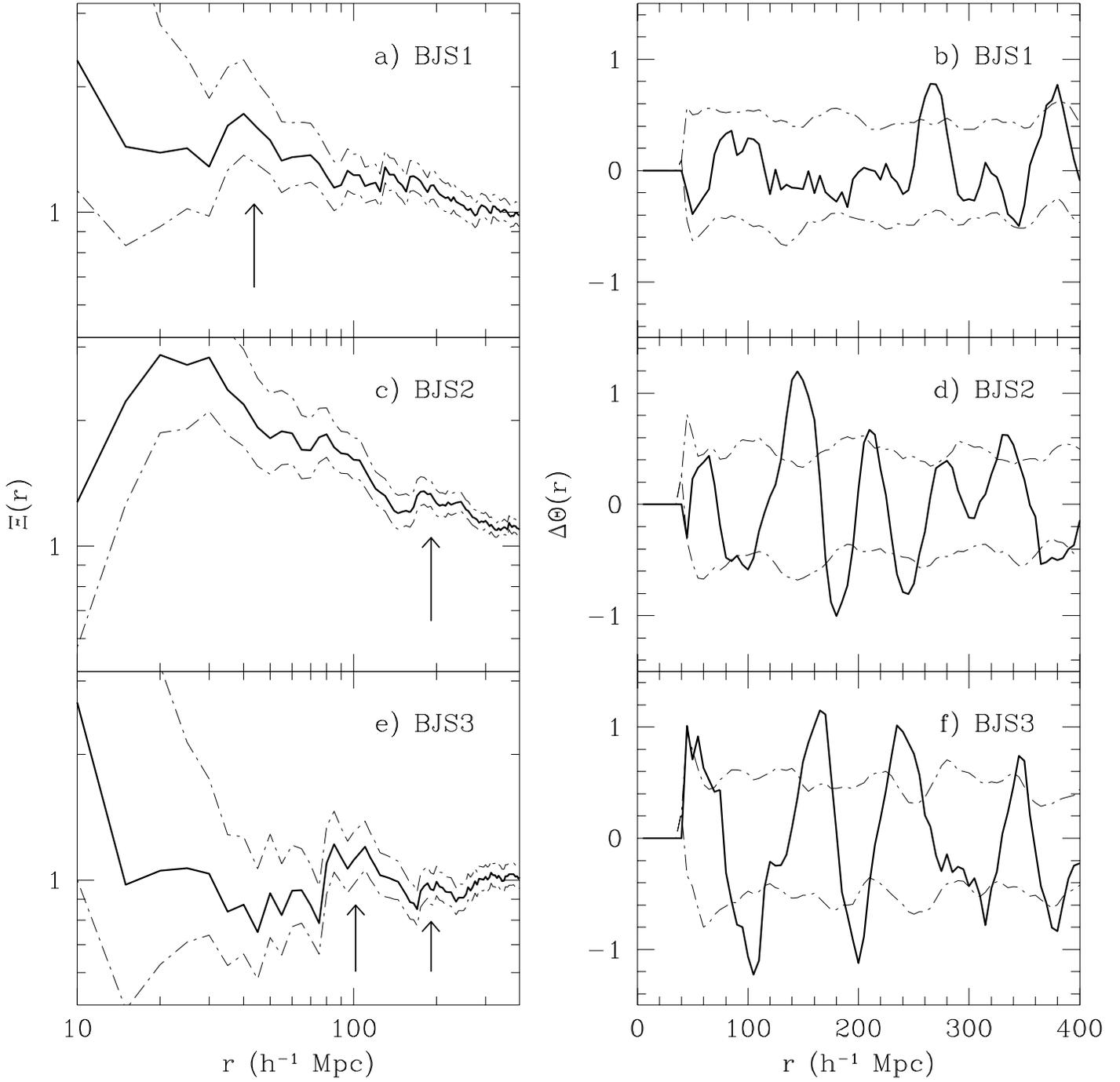
{\bf Figure 2.} The integral two-point spatial correlation function
$\overline{\xi}(r)$ for the BJS1, BJS2, and BJS3 samples (panels a,c,
and e, respectively)
and its smoothed second derivative $\Delta\Theta(r)$ (panels b,d, and f). 
The structure of the plots is the same as for Fig.1.

\begin{thebibliography}{99}

\bibitem[Andreani \& Cristiani 1991]{AC92} 
          Andreani, P., Cristiani, S., 1992, ApJ, 398, L13
\bibitem[Boyle et al. 1991]{BJS}
          Boyle, B.J., Jones, L.R., Shanks, T., 1991,
          MNRAS, 251, 482 (BJS)
\bibitem{} Clowes, R.G., Campusano, L.E., 1991a, 
            MNRAS, 249, 218
\bibitem{} Clowes, R.G., Campusano, L.E., 1991b, 
            in {\em The Space Distribution
            of Quasars, ASP Conf. Ser.},{\bf 21}, D.Crampton (ed.), 248.
\bibitem{} Crampton, D., Cowley, A.P., Hartwick, F.D.A., 1987,
            ApJ, 314, 129
\bibitem{} Crampton, D., Cowley, A.P., Hartwick, F.D.A., 1989,
            ApJ, 345, 59
\bibitem{} Deng, Z., Xia, X., Fang, L.-Z., 1994, 
           ApJ, 431, 506
\bibitem{} Dinshaw, N., Impey, C.D., 1996, ApJ, 458, 73
\bibitem[Einasto et al. 1984]{FOF}
             Einasto, J., Klypin, A.A., Saar, E., Shandarin, S.F., 1984,
            MNRAS, 206, 529
\bibitem{} Graham, M.J., Clowes, R.G., Campusano, L.E., 1995,
            MNRAS, 275, 790
\bibitem[Hall et al. 1996]{}
           Hall, P.B., Osmer, P.S., Green, R.F., Porter, A.C., 
           and Warren, S.J., 1996, ApJ, 462, 614
\bibitem{} Jakobsen, P., Perryman, M.A.C., Ulrich, M.H., Maccheto, F., 
           Di Serego Alighieri, S., 1986, 
            ApJ, 303, L27
\bibitem{} Jakobsen, P., Perryman, M.A.C., 1992, 
            ApJ, 392, 432
\bibitem[Iovino et al. 1991]{Iovinoetal91} 
            Iovino, A., Shaver, P., Cristiani, S., 1991, in
            {\em The Space Distribution of Quasars, ASP Conf. Ser.}, 
            {\bf 21},
            D. Crampton (ed.), 202.
\bibitem[Komberg \& Lukash 1994]{KL}
           Komberg, B.V., Lukash, V.N., 1994, MNRAS, 269, 277
\bibitem[Komberg et al. 1994]{paperI} 
           Komberg, B.V., Kravtsov, A.V., Lukash, V.N., 1994, 
           A\&A, 286, L19
\bibitem[Komberg et al. 1996]{paperII} 
           Komberg, B.V., Kravtsov, A.V., Lukash, V.N., 1996,
           MNRAS, 282, 713 (preprint {\tt astro-ph/9602090})
\bibitem{} Koo, D.C., Kron, R.G., 1988, ApJ,
           325, 92 (KK)
\bibitem{} Lespine, Y., Petitjean, P., 1996, A\&A, in press,
           preprint {\tt astro-ph/9606037}
\bibitem{} Mo, H.J., Deng, Z.G., Xia, X.Y., Schiller, P., 
           B\"{o}rner, 1992, A\&A, 257, 1
\bibitem[Mo \& Fang 1993]{MF93} 
           Mo,H., Fang, L.-Z., 1993, ApJ {\ } 410, 493
\bibitem{} Sargent, W.L.W., Steidel, C.C., 1987, 
           ApJ, 322, 142
\bibitem[Shanks \& Boyle 1994]{SB94} 
           Shanks, T., Boyle, B.J., 1994, MNRAS, 271, 753
\bibitem{} Webster, A., 1982, MNRAS, 199, 683
\bibitem[Williger et al. 1996]{Williger96}
           Williger, G.M., Hazard, C., Baldwin, J.A., and 
           McMahon, R.G. 1996, ApJS, 104, 145
\bibitem{} Zitelli, V., Mignoli, M., Zamorani, G., Marano, B.,
           Boyle, B.J., 1992, MNRAS, 256, 349 ((ZM)$^2$B)


 \end{thebibliography}
\end{document}